# Towards SAMBA: Segment Anything Model for Brain Tumor Segmentation in Sub-Sharan African Populations

Mohannad Barakat[1,2*], Noha Magdy[1,2*], Jjuuko George William[3], Ethel Phiri[4], Raymond Confidence[5,6], Dong Zhang[7] and Udunna C Anazodo[5,6, 7,8,9,10]

[1]Department of Engineering and Applied Sciences, Nile University, Giza, Egypt
[2]Department of Computer Science, Friedrich-Alexander-Universität, Erlangen, Germany
[3]Makerere University, Kampala, Uganda
[4]Malawi University of Science and Technology, Malawi
[5] Medical Artificial Intelligence Laboratory (MAI Lab), Lagos, Nigeria
[6]Lawson Health Research Institute, London, Ontario, Canada.
[7]Department of Electrical and Computer Engineering, University of British Columbia, Vancouver, Canada
[8]Montreal Neurological Institute, McGill University, Montréal, Canada.
[9]Department of Medicine, University of Cape Town, South Africa
[10]Department of Clinical & Radiation Oncology, University of Cape Town, South Africa

mohannad.barakat@fau.de

**Abstract.** Gliomas, the most prevalent primary brain tumors, require precise segmentation for diagnosis and treatment planning. However, this task poses significant challenges, particularly in the African population, were limited access to high-quality imaging data hampers algorithm performance. In this study, we propose an innovative approach combining the Segment Anything Model (SAM) and a voting network for multi-modal glioma segmentation. By fine-tuning SAM with bounding box-guided prompts (SAMBA), we adapt the model to the complexities of African datasets. Our ensemble strategy, utilizing multiple modalities and views, produces a robust consensus segmentation, addressing the intra-tumoral heterogeneity. Although the low quality of scans presents difficulties, our methodology has the potential to profoundly impact clinical practice in resource-limited settings such as Africa, improving treatment decisions and advancing neuro-oncology research. Furthermore, successful application to other brain tumor types and lesions in the future holds promise for a broader transformation in neurological imaging, improving healthcare outcomes across all settings. This study was conducted on the Brain Tumor Segmentation (BraTS) Challenge Africa (BraTS-Africa) dataset, which provides a valuable resource for addressing challenges specific to resource-limited settings, particularly the African population and facilitating the development of effective and more generalizable segmentation algorithms. To illustrate our approach's potential, our experiments on the BraTS-Africa dataset yielded compelling results, with SAM attaining a Dice coefficient of 86.6 for binary segmentation and 60.4 for multi-class segmentation. These outcomes serve as a testament to the promise and effectiveness of our proposed framework.

---

[*] Equal contribution

# 1  Introduction

Brain tumors represent a significant global health challenge, affecting millions of lives each year [1]. Among these tumors, gliomas being the most prevalent primary brain tumor is characterized by their heterogeneous and infiltrative nature, making diagnosis and treatment challenging [1]. The complexity of gliomas stems from their morphological and biological variations, leading to intricate sub-regions with distinct characteristics. Precise segmentation of these sub-regions, such as the active tumor core, peritumoral edema, and enhancing tumor regions, is crucial for understanding the tumor's behavior and guiding personalized treatment strategies [2]. However, achieving precise glioma segmentation remains a challenging task [3], particularly in resource-limited settings where access to high quality advanced brain imaging tools and skilled personnel to manually analyze high volume of imaging data, remain scarce [4]. Specifically, for Sub-Saharan African populations, accurate segmentation is critical because of the usual delayed disease presentation and the high propensity for comorbidities such as infectious disease. This leads to misdiagnosis and worse outcomes [5][6]. Thus, this study aims to provide an adaptive and robust methodology that can improve the accuracy of glioma segmentation in low-resourced settings and pave the way for advancements in neuro-oncology research.

In recent years, various approaches have been explored for the segmentation of brain tumor data [8], each aiming to achieve improved performance. These include learning frameworks for automatic detection of tumor boundaries, such as DeepSeg [9], nnU-Net [10], and DeepSCAN [11], as well as approaches such as Swin UNETR [12] based on vision transformers. Another promising method trained on multiple U-net-like neural networks with deep supervision and stochastic weight averaging, produce segment brain tumor subregions by assembling models from different training pipelines [13]. More recently, the Segment Anything Model (SAM) [14] was introduced as a pioneering image segmentation solution, known for its exceptional ability to generate high-quality object masks. Whether prompted by points or boxes, SAM effortlessly produces accurate masks for diverse objects within images [14]. Trained on a vast dataset of 11 million images and 1.1 billion masks, SAM's revolutionary zero-shot capabilities set it apart from conventional methods, making it indispensable for various segmentation tasks. [14]. The adoption of SAM in the medical field has shown potential, particularly when fine-tuned [15] [16]. By fine-tuning SAM [14], we adapt the model to focus on the region of interest within the brain, making it better equipped to handle the complexities of African datasets. This targeted fine-tuning process enables SAM to extract relevant features from the limited and potentially noisy imaging data, enhancing the accuracy of glioma segmentation. The integration of multiple imaging modalities, including FLAIR, T1-weighted, T2-weighted, and T1-enhanced, is vital for gaining a comprehensive understanding of the glioma's characteristics. To this end, we utilize a voting network ensemble strategy, which combines individual segmentations from SAM generated using different modalities and image views. This ensemble approach aims to mitigate the uncertainties and artifacts present in individual modalities, ultimately providing a more robust consensus segmentation.

Since 2012, the Brain Tumor Segmentation (BraTS) Challenge has offered open MRI training data, annotations, and model evaluation metrics, catalyzing machine learning (ML) progress in glioma diagnosis [4]. Uncertainty persists about whether advanced ML methods developed from BraTS data can be applied in Sub-Saharan clinical settings given their unique challenges including the limited number of annotated cases for model training and validation, and the lower resolution of acquired MRI. Here, we leveraged the recently introduced BraTS-Africa dataset [6] to explore the potential of fine-tuning SAM to improve the accuracy of glioma segmentation and provide a viable solution to overcome these unique challenges.

## 2      Methodology

### 2.1      The Dataset

The dataset comprised of 60 (45 training, 15 validation) pre-operative adult glioma cases from the MICCAI-CAMERA-Lacuna Fund BraTS-Africa 2023 Challenge data [6] and 250 (200 training, 50 validation) adult glioma cases from the BraTS 2021 Challenge data [6] [7]. Each case included routine multi-parametric MRI T1-weighted, T2-weighted, T2-FLAIR and T1-post-contrast enhanced (T1CE) scans, meticulously annotated by experienced neuroradiologists for training, validation, and testing (Figure 1). [6] [7]. The annotated sub-regions are the "Enhancing tumor" (ET), "Non-enhancing tumor core" (NETC), and "Surrounding non-enhancing FLAIR hyperintensity" (SNFH). The ET are areas with increased T1 signal on postcontrast images, while the NETC comprises of the non-enhancing tumor core regions, including necrosis and cystic changes, and the SNFH refers to FLAIR signal abnormality surrounding the tumor but not part of the tumor core [6].

### 2.2      The SAM Model

The Segment Anything Model (SAM) incorporates a promptable design that facilitates interactive specification of the target area for image segmentation with minimal human intervention. SAM's design includes an image encoder, a prompt encoder, and a lightweight decoder for generating segmentation masks (Figure 2). Drawing inspiration from chat-based Large Language Models, SAM allows users to provide prompts to guide the segmentation process effectively [13]. SAM supports three distinct types of prompts:
    1.   Point Prompt: Users select a point in the image to define the target area.

2. Bounding Box Prompt: Users draw a bounding box around the object to be segmented.
3. Rough Mask Prompt: Users manually draw a basic mask outlining the target object.

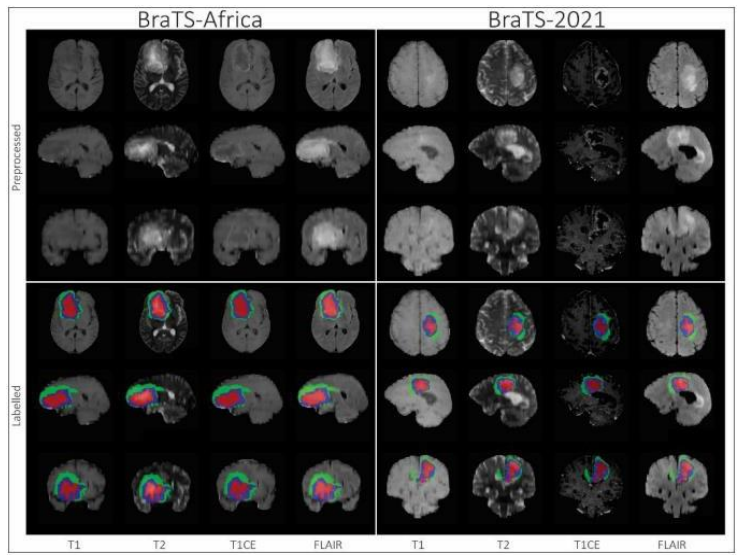

**Figure 1.** Clinical standard brain MRI typically acquired in Sub-Saharan African Populations illustrated in a glioma patient (left) shows the conventional lower image resolution compared to the BraTS 2021 data (right). Adapted from [6]

In addition to those prompts SAM has a not yet implemented text prompt. By utilizing these prompts, SAM becomes flexible, adaptable, and capable of handling novel objects not explicitly present in its training data. This feature empowers SAM as a powerful and versatile image segmentation model[*].

Experiments were conducted in this research with various prompts to guide SAM for glioma segmentation. However, implementing the "points" prompt, lead to significant variations in the generated masks with minor pixel changes, due to its high sensitivity (see Equation. 1). To clarify, selecting a positive point within the tumor as a prompt, and subsequently choosing another point within the tumor with a slight pixel shift, may result in the generation of two entirely disparate masks which makes point prompt very sensitive to slight changes in the input points. Similarly, using the "bounding box" prompt around the entire brain image proved ineffective, as SAM struggled to discern the tumor area amidst the extensive brain region. As a result, we decided to focus on the "bounding box" prompt around the tumor area, as it had shown promising initial results even before fine- tuning.

---

[*] Codes are available at:
https://github.com/CAMERAMRI/SPARK2023/tree/main/SPARK_SAMBA

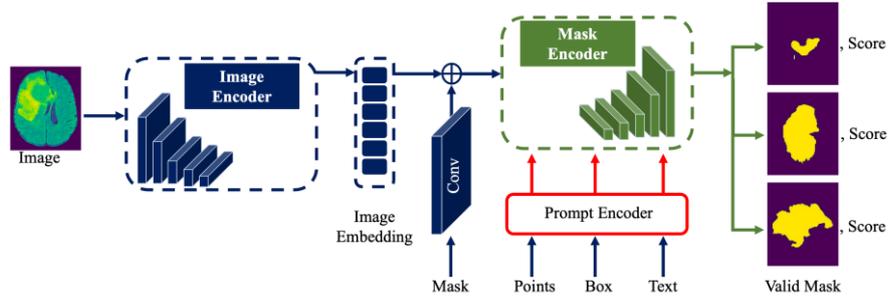

**Figure 2.** SAM model architecture

Table.1 Summarizes the prompts experimentation. To obtain the desired bounding box, we fine-tuned a YOLO v8 localizer [17] for 150 epochs. The YOLO localizer, based on the You Only Look Once algorithm [17], excels in real-time object detection by simultaneously predicting bounding boxes and class probabilities. By using these localized bounding boxes as prompts, SAM's segmentation was targeted, leading to improved accuracy and robustness in glioma segmentations.

*Assume prompt* $p_1 = (x, y)$ *with mask* $m_1$
*prompt* $p_2 = (x + 1, y)$ *with mask* $m_2$

*A non-sensitive prompt means that* $IoU(m_1, m_2) \approx 1$ (Equation 1)

**Table 1.** advantages and disadvantages for different modes of SAM

| Prompt | Explanation | Dis/Advantage | Status |
|---|---|---|---|
| **Point** | Point(s) inside the lesion. | - Easy to compute.<br>- SAM is not stable with this prompt. | Needs investigation |
| **Box** | Bounding box around lesion. | - Easy to compute.<br>- SAM is stable with this prompt | **This paper** |
| **Mask** | Initial binary mask covering part of or the whole lesion. | - Hard to get. | Future work |
| **Text** | Input text describing the type of lesion to segment. | - Allows SAM to segment all lesion.<br>- Hardest to train | Future work |

### 2.3 SAMBA: Finetuning SAM

This study proposes two distinct approaches for fine-tuning SAM using the bounding box prompt mode to address the task of accurate glioma segmentation (SAMBA). In both approaches, a localizer is employed to generate bounding boxes encompassing the glioma regions, providing spatial information to guide SAM's segmentation. In the first approach, SAM's image and prompt encoders remain frozen, while the decoder is specifically fine-tuned for binary segmentation of gliomas without specifying the three classes. Subsequently, a compact voting network is introduced to amalgamate the binary segmentation outputs from SAM across various modalities and image views, yielding a final three-class glioma segmentation that effectively captures intra-tumoral heterogeneity. Figure 3 shows the SAMBA architecture for this approach.

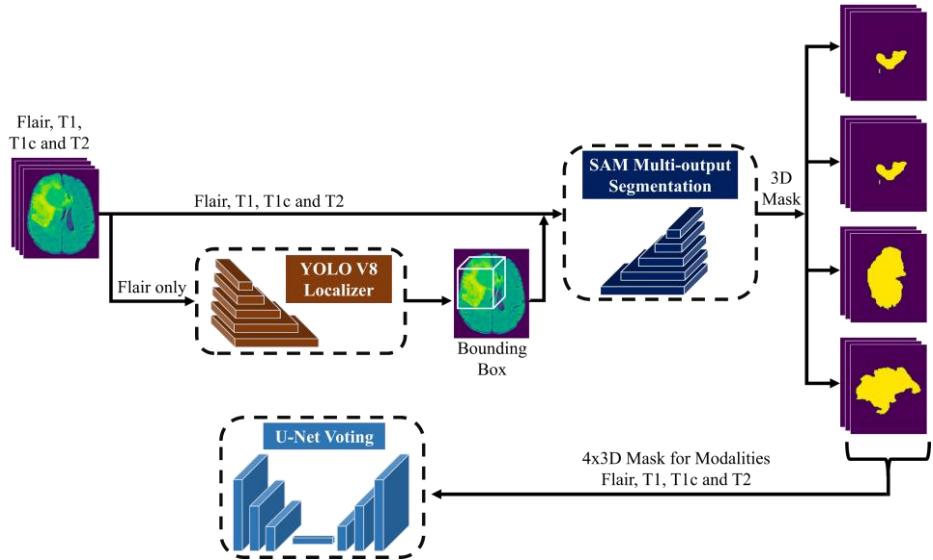

**Figure 3.** The architecture for SAM decoder fine-tuning for binary segmentation (SAMBA).

The second approach also involves freezing the encoders and fine-tuning the decoder, but this time, it is tailored to directly output the three designated segmentation classes. Optionally, a voting network is integrated to further enhance the results by leveraging the encoder's output to capture finer details not entirely captured by SAM. Figure 4 shows the architecture of SAM multi class finetuning (SAMBA-mc).

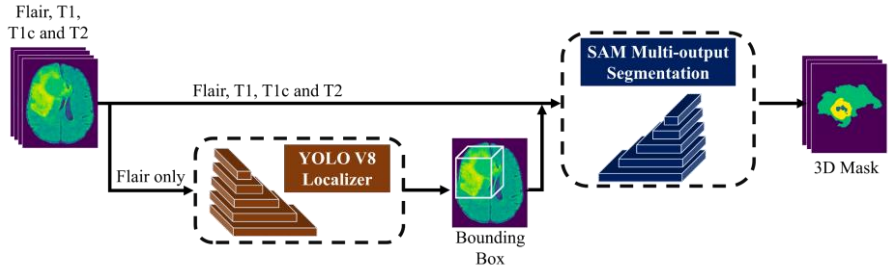

**Figure 4.** The architecture for SAM decoder fine-tuning for multi-class segmentation (SAMBA-mc).

In the binary segmentation approach (SAMBA), SAM was trained for 15 epochs, whereas in the multi-class segmentation approach (SAMBA-mc), SAM was trained for 50 epochs. The shorter training duration for binary segmentation was sufficient to achieve satisfactory results, as it involved a simpler task of distinguishing the presence or absence of the tumor. On the other hand, the multi-class segmentation task required a more extended training period to accurately classify the tumor into three distinct classes: edema, enhancing tumor, and non-enhancing tumor. The different training epochs were tailored to the complexity and requirements of each segmentation task, optimizing the performance of SAM in both scenarios.

### 2.4    The Voting Network

The voting network used in our study is a lightweight 3D U-Net with approximately 200,000 parameters. It serves two primary purposes: Firstly, it addresses SAM's limitations in capturing fine details by refining the segmentation results. Secondly, it performs a crucial voting process by aggregating the binary segmentation outputs from SAM across different modalities (flair, t1, t2, t1 enhanced). By combining the segmented outputs through voting, the voting network produces the final three-class segmentation. The incorporation of the voting network significantly enhances the robustness and accuracy of the glioma segmentation, addressing both SAM's limitations and the complexities posed by multi-modal data.

## 3    Results

We observed that the fine-tuning of SAM's decoder for binary segmentation (SAMBA) resulted in a Dice coefficient of 63.3%, which represented a decrease in accuracy compared to the 72.2% achieved on higher-quality images from the BraTS 2021 dataset when trained using a bounding box around the whole brain. However, when SAM was fine-tuned using a bounding box around the tumor region obtained from the localizer, the accuracy significantly improved to 84.6% on the BraTS-Africa data and 89.4% on the better-quality data from BraTS 2021 (Figure 5). Notably, SAM without fine-tuning achieved a Dice coefficient of 73.6% when using a bounding box around the tumor on the BraTS 2021 dataset. This highlights the efficacy of SAM's promptable design, as it performed well even without specific fine-tuning on the high-quality dataset, further demonstrating its potential for accurate glioma segmentation. Figure 6 shows the improvement of SAM before and after fine-tuning.

For the multi-class segmentation task, SAM was trained on the African data, and its Dice coefficient reached the value of 60.4%. Integrating YOLO into the pipeline decreased the Dice coefficient value, impacting segmentation results. SAM performed well initially using ground truth bounding boxes, but when tested with YOLO- generated bounding boxes, the Dice coefficient decreased. Enhancing YOLO's performance could improve the overall Dice coefficient, highlighting the need to refine the interaction between YOLO and SAM for better segmentation accuracy. Table. 2 summarize SAM results. These findings suggest that fine-tuning SAM using the localized bounding box around the tumor region significantly improved segmentation accuracy, particularly in the African dataset, which is characterized by lower image quality. The results demonstrate the potential of SAM's adaptability to different data qualities and its ability to produce accurate segmentations with appropriate guidance.

**Table 2.** Dice scores for finetuned SAM (SAMBA) model on both BraTS-Africa data and BraTS 2021 data on binary and multi-class segmentation.

|  | BraTS-Africa (Dice) | BraTS 2021 (Dice) |
| --- | --- | --- |
| Binary, full brain box | 63.3 % | 72.2 % |
| Binary, tumor bounding box (without YOLO) | 84.6 % | 89.4 % |
| Binary, tumor bounding box (with YOLO) | 33.7% |  |
| Multi class tumor bounding box (without YOLO) | 60.4% (mean over ET, TC, WT) | - |
| Multi class tumor bounding box (with YOLO) | 12.8% (ET), 16.8% (TC), 50.9% (WT) | - |

The YOLO v8 localization network achieved a high accuracy of 96.7% in correctly differentiating between images with bounding boxes (i.e., slices containing tumors) and background images (i.e., slices without tumors). For the bounding boxes generated by the model, the box loss was 1.24, indicating accurate localization, with an average box confidence of 87%. Some examples of the results shown in Figure 6. These improved results demonstrate the effectiveness of the YOLO v8 localizer in accurately detecting and localizing tumor regions within the brain images, providing valuable bounding boxes to guide the glioma segmentation process in our research.

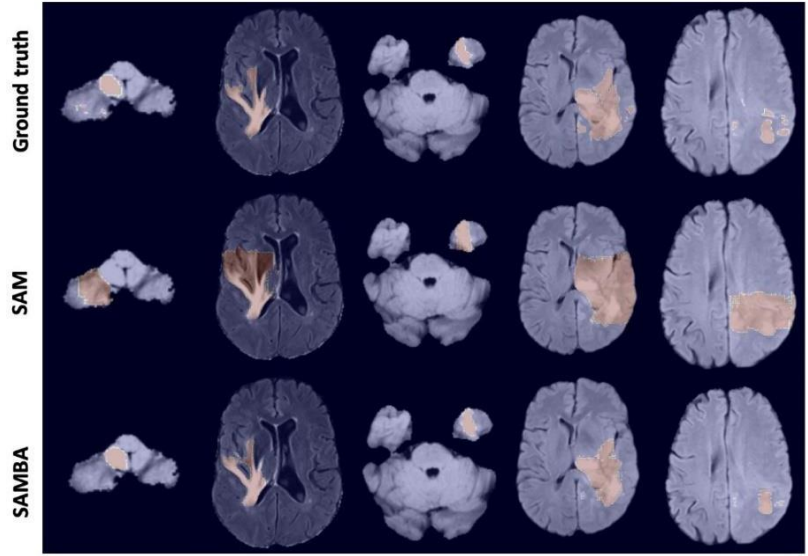

**Figure 5.** An example of SAM results illustrating improvement before and after fine-tuning for binary segmentation in BraTS-Africa. The results also show the limitation of SAM on capturing fine details.

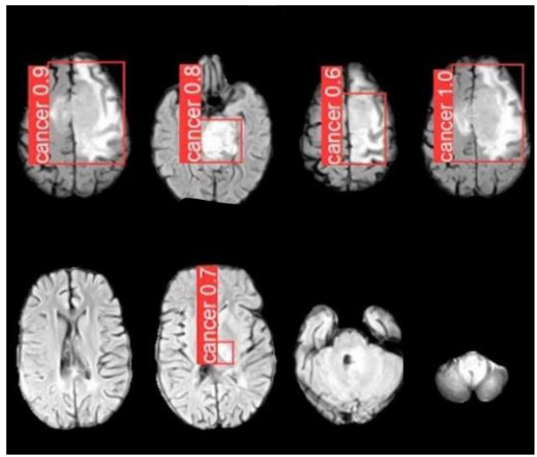

**Figure 6.** Example of the localization output results indicated by box confidence score for the BRATS Africa dataset. The box outlined in red and the confidence of the model on detecting the class "cancer" are overlaid on the figured where the tumor was identified.

With the constraint imposed by YOLO, the overall Dice coefficient reached approximately 43%, showcasing the voting network's valuable contribution to mitigating the impact of this limitation.

## 4    Discussion

This study demonstrated the effectiveness of SAM in glioma segmentation, particularly when guided by bounding boxes around the tumor region. Fine-tuning SAM (SAMBA) using the localized bounding boxes significantly improved segmentation accuracy, achieving Dice coefficients of 84.6% on the BraTS-Africa dataset and 89% on the higher-quality BraTS 2021 dataset. Notably, SAM performed well without fine-tuning, achieving a Dice coefficient of 73% on BraTS 2021 data with a bounding box around the tumor. SAM showed that it could reach high DICE coefficient when fintuning on limited data (only 45 cases), an important finding for low-resourced settings which tend to have fewer training and validation datasets.

Using SAM with multi-class segmentation does not yield improved results. That possibly due to the difference between the purpose of the multi-class segmentation in SAM and SAMBA. In SAM each layer represented a sub-object to the inferior layer. While this remains the same for SAMBA, SAMBA has multiple outputs where it should only have one or two outputs not three. Fine tuning SAM with a fixed prompt around the brain didn't yield good DICE score. However, this was for fine tuning the decoder only. Fine tuning the encoder and decoder together might give better results and will be investigated in future work.

The incorporation of a lightweight U-Net voting network further enhanced segmentation results, addressing SAM's limitations in capturing fine details. The voting network effectively combined binary segmentation outputs from different modalities and views, producing robust three-class glioma segmentations. Moreover, the utilization of the YOLO v8 localization network for accurate tumor detection proved instrumental in providing reliable bounding boxes to guide SAM's segmentation process. This approach proved particularly valuable for datasets with lower image quality, such as the African dataset, where access to high-quality imaging data is limited. Although YOLO v8 made SAMBA possible by providing prompts, it clearly degraded the overall loss by introducing some false positives and false negatives.

Overall, our findings demonstrate the potential clinical impact of SAM for brain tumor imaging and treatment planning, especially in regions with limited imaging resources. The adaptability and versatility of SAM make it a valuable tool for accurate and efficient lesion segmentation, paving the way for extension of SAMBA to brain metastasis and other types of brain lesions to further improved diagnostic and treatment decisions, especially in cases with complex brain tumor features. As SAMBA continues to evolve, future applications hold promise for broader transformations in neurological imaging to advance healthcare outcomes.

## 5   Future Work

For future work, we propose to explore fine-tuning the entire SAM, including both the encoder and decoder, using the LoRA fine-tuning technique [18]. Given the substantial size of the encoder, LoRA fine-tuning offers a more efficient approach to update the model's parameters while preserving its learned knowledge. We anticipate that fine-tuning the full SAM will yield significantly improved segmentation results com- pared to the current approach. Additionally, we plan to develop a specialized version of SAM tailored specifically for brain tumor segmentation, capable of generalizing beyond gliomas to other brain tumor types. Examples of these brain tumor types could include meningiomas, metastatic tumors, pituitary adenomas, and others. By incorporating diverse tumor types in the training data, the fine-tuned SAM encoder will be able to learn the distinct characteristics and variations across these tumor types, ultimately leading to highly accurate segmentations for each specific tumor type.

## 6   Acknowledgment

The authors would like to thank the all the faculty and instructors of the Sprint AI Training for African Medical Imaging Knowledge Translation (SPARK) Academy 2023 summer school on deep learning in medical imaging for providing insightful background knowledge on brain tumors that informed the research presented here. The authors would also like to thank Linshan Liu for administrative assistance in supporting the SPARK Academy training and capacity building activities which the authors immensely benefited from. The authors acknowledge the computational infrastructure support from the Digital Research Alliance of Canada (The Alliance) and knowledge translation support from the McGill University Doctoral Internship Program through student exchange program for the SPARK Academy. The authors are grateful to McMedHacks for providing foundational information on python programming for medical image analysis as part of the 2023 SPARK Academy program. This research was funded by the Lacuna Fund for Health and Equity (PI: Udunna Anazodo, grant number 0508-S-001) and National Science and Engineering Research Council of Canada (NSERC) Discovery Launch Supplement (PI: Udunna Anazodo, grant number DGECR-2022-00136).